# The link between countries' economic and scientific wealth has a complex dependence on technological activity and research policy

Alonso Rodríguez-Navarro[a,b]*, Ricardo Brito[b]

[a] *Departamento de Biotecnología-Biología Vegetal, Universidad Politécnica de Madrid, Avenida Puerta de Hierro 2, 28040, Madrid, Spain*

[b] *Departamento de Estructura de la Materia, Física Térmica y Electrónica and GISC, Universidad Complutense de Madrid, Plaza de las Ciencias 3, 28040, Madrid, Spain*

*\* Corresponding author: e-mail address: alonso.rodriguez@upm.es*

**Abstract**

We studied the research performance of 69 countries by considering two different types of new knowledge: *incremental* (*normal*) and *fundamental* (*radical*). In principle, these two types of new knowledge should be assessed at two very different levels of citations, but we demonstrate that a simpler assessment can be performed based on the total number of papers (P) and the ratio of the number of papers in the global top 10% of most cited papers divided to the total number of papers ($P_{top\ 10\%}/P$). P represents the quantity, whereas the $P_{top\ 10\%}/P$ ratio represents the efficiency. In ideal countries, P and the $P_{top\ 10\%}/P$ ratio are linked to the gross domestic product (GDP) and GDP the per capita, respectively. Only countries with high $P_{top\ 10\%}/P$ ratios participate actively in the creation of *fundamental* new knowledge and have Noble laureates. In real countries, the link between economic and scientific wealth can be modified by the technological activity and the research policy. We discuss how technological activity may decrease the $P_{top\ 10\%}/P$ ratio while only slightly affecting the capacity to create *fundamental* new knowledge; in such countries, many papers may report *incremental* innovations that do not drive the advancement of knowledge. Japan is the clearest example of this, although there are many less extreme examples. Independently of technological activity, research policy has a strong influence on the $P_{top\ 10\%}/P$ ratio, which may be higher or lower than expected from the GDP per capita depending on the success of the research policy.

*Key words: research efficiency; economic wealth; fundamental knew knowledge; incremental new knowledge; research assessment*

## 1. Introduction

It is widely accepted that research, development, and innovation (RDI) play a central role in the economic development of countries. Indeed, dramatic technological changes generated by RDI can be observed over the period of a few years in the devices that we use every day.

Numerous publications have linked RDI with economic progress, and the Organisation for Economic Co-operation and Development (OECD) and diverse authors have coined the terms "knowledge-based economy," "new economy," and "economics of science" to described this (e.g., Godin 2004, 2006; Harris 2001; Dasgupta and David 1994; Coccia 2018).

Although these concepts apply to all countries and most high- and middle-income countries make substantial investments in RDI and publish many scientific papers, the success of their research varies greatly. For example, Nobel laureates in sciences are concentrated in very few countries (Schlagberger et al. 2016), which suggests that, in the others, independently of how many papers are published, their research never reaches scientific achievements at the Nobel level. Even in countries with Nobel laureates in the natural sciences, which one might suppose to be the most research advanced in research terms, the number of papers that have to be published to be awarded a Nobel Prize varies enormously, and this variability becomes still greater if research institutions are included in the comparison (Rodríguez-Navarro 2011). These observations lead to the conclusion that, regarding the real contribution of countries to the advancement of science, their efficiency in producing these advancements, which is independent of size, might be as important as the total amount of research that they perform.

Research efficiency has been investigated (Bonaccorsi and Daraio 2004; Daraio 2019) and Sandström and Van den Besselaar (2018) studied the potential causes of differences in research efficiency among countries. However, before drilling down into the details of what makes a research system more or less efficient, the economic factor must also be considered. Many studies have demonstrated that scientific wealth depends on economic wealth (Cimini et al. 2014; Allik et al. 2020; Cole and Phelan 1999; Docampo and Bessoule 2019; Gantman 2012; King 2004; May 1997; Mueller 2016; Rousseau and Rousseau 1998; Rahman and Fukui 2003), which suggests that countries with medium or low GDP per capita might contribute very little to scientific progress. Such countries may have insufficient economic resources to fund and maintain a competitive research system or to develop the efficient functional structures required to perform research efficiently and compete successfully with richer countries in the advancement of science.

However, the economy may not be the only factor sustaining successful research in high-income countries. For example, the proportion of countries' papers in the global top 10% or 1% of most cited papers (Bornmann et al. 2015) varies notable even among those that apparently do not vary greatly in economic level. Moreover, in the European Union (EU), although the procedures for selecting researchers and awarding funds to projects applied by the European Research Council are similar for all countries, the successes of the awarded projects varies depending on the country in which the research is carried out, including among countries with similar economies (Rodríguez-Navarro and Brito 2020b).

All this suggests that the efficiency of countries in contributing to scientific progress depends on their economic wealth, but also on other factors that can be subsumed into the concept of research policy. The question that thus arises is about the independent roles that economic wealth and research policy play in research efficiency. However, the answer to this question is anything but simple without prior agreement on how to measure scientific progress, which is still lacking.

According to the Frascati Manual (OECD 2002, pp. 30 and 34), research is "*creative work undertaken on a systematic basis in order to increase the stock of knowledge,*" including the presence "*of an appreciable element of novelty and the resolution of scientific and/or technological uncertainty.*" Although these definitions are clear, the procedure required to measure such contributions to new knowledge creation is not as clear because it should also include the relevance of the new knowledge, which is not currently included. It is obvious that not all new knowledge is equally relevant to science and society. Therefore, each of the above-cited studies about research and economic wealth applies its own indicator without empirically demonstrating that it is the most appropriate, or even that a single indicator is sufficient. For assessment purposes, the National Science Board of the USA reports countries' share of papers in four top percentiles of the globally most cited papers: 50, 10, 5, and 1. However, they state that "*the relative impact of an economy's S&E research can be compared through the representation of its articles among the world's top 1% of cited articles*" (National Science Board 2020, p. 12), without justifying the use of this percentile rather than the top 0.1% or 10%.

Another illustrative example is the adjective that it is used to qualify research with the greatest relevance: "excellent." Even OECD publications about research use the terms "excellence" or "excellent," omitting a numerical or level definition of where excellence might start (Rodríguez-Navarro and Brito 2018, p. 719).

## 2. Measuring contributions to new scientific knowledge creation

This brief introduction above highlights that any study that attempts to separate the effects of economic wealth and research policy on knew knowledge creation might be controversial without a prior agreement regarding how to measure new knowledge creation.

There are well-established methods for research assessment based on citations (van Raan 2019; Waltman and van Eck 2019), but the lack of agreement about the most appropriate indicator to measure contributions to new knowledge creation is not methodological but rather arises because the relevance of the new knowledge that should be measure remains open to interpretation.

The widely known study by Khun (1970) about scientific revolutions presents *normal science* as the product of the day-to-day work of researchers; in contrast, *revolutionary science* is an infrequent product of research that has a much greater effect on the progress of science. These ideas have been extended globally and presented in multiple forms. For example, Chapter 1 of the textbook of (Chen 2013, p. 1) begins with the statement that "*Scientific knowledge changes all the time. Most of the changes are incremental, but some are revolutionary and fundamental*". Therefore, when trying to measure research efficiency, the first question that arises is whether this refers to the creation of *fundamental* (*revolutionary*) or *incremental* (*normal*) new knowledge.

The problem of quantifying *fundamental* new knowledge using bibliometric methods lies in the infrequency with which such papers are published. If we use top percentiles for research assessment (Bornmann et al. 2013), there is no fixed top percentile that separates *fundamental* from *incremental* new knowledge. However, researchers in many fields would probably agree that less than 1 paper per 1,000 is truly *revolutionary*. This low proportion implies that it would be necessary to quantify the number of papers published by a country that make it into the global top 0.1% or perhaps even the top 0.01% of cited papers. It is obvious that such values for many countries would be too low to be counted with minimal statistical reliability. However, this issue may be addressed by calculating the probability or expected frequency of such papers (Rodríguez-Navarro and Brito 2019).

Meanwhile, *incremental* knowledge provides support for *fundamental* knowledge, and from a scientific point of view, all scientific publications should be considered to report *incremental* new knowledge. In essence, there are no conceptual reasons for exclusions, except for the insignificant number of papers that report *fundamental* knowledge.

According to this reasoning, the independent effects of economic wealth and research policy on the scientific wealth of countries cannot be studied using a single citation indicator. At least, the total number of papers (henceforth P) and the number of papers that are included in a certain global top percentile (henceforth $P_{top\ x\%}$) should be used.

Only two parameters of this type are needed because the percentile distribution of papers based on citation counts follows a power law (Brito and Rodríguez-Navarro 2018); for evaluative purposes, the exponent or scaling parameter of the power law is used to calculate its derivative $e_p$ (Rodríguez-Navarro and Brito 2018). The $e_p$ constant equals 0.1 raised to a power that is the exponent of the power law, which is calculated by fitting a power law to the number of papers in six or more top percentiles. Using the $e_p$ constant the following equations apply (Rodríguez-Navarro and Brito 2019):

The probability that of a country's paper will appear in $P_{top\ x\%}$ is $\quad e_p^{(2 - \lg x)} \quad$ (1)

The expected frequency of such a paper is $\quad P \cdot e_p^{(2 - \lg x)} \quad$ (2)

And thus $\quad e_p = P_{top\ 10\%}/P \quad$ (3)

According to Eq. 3, when only $P_{top\ 10\%}$ and P are known, $P_{top\ 10\%}/P$ is a proxy of $e_p$.

## 3. Incremental innovations and possible failure of some citation metrics

This reasoning about papers that report *incremental* and *fundamental* knowledge assumes that the aim of the publications considered is to achieve the progress of scientific knowledge.

However, this is not always true because a significant amount of global research is addressed toward improving technological processes and products rather than the advancement of science per se. Although the results of such research are patented, a significant amount is also published in scientific journals. This issue is highly relevant because, in advanced countries, the volume of such research may be very high. In fact, most of the improvements that continuously appear in the devices that we use every day are *incremental* innovations; if such new knowledge is published, it should not be included in the value of P used in Eq. 2 and 3.

For example, consider the field of rechargeable lithium batteries, which has become very important because of their use in electronic devices and electric vehicles (Goodenough and Kim 2010). The improvements in this type of batteries over the last 30–40 years have been spectacular, and the number of papers in this field exceeds 200,000. *Fundamental* innovations have been achieved in this area, as illustrated by the three Nobel Prizes in chemistry awarded in 2019. However, as mentioned above, very few papers in this field report *fundamental* research. One such paper reporting an innovation regarding the positive electrode of the battery (Padhi et al. 1997) lies in the top 0.01% of all papers published in the field (up to and including its year of publication) or in the field of chemistry in the WoS (SU=chemistry) in its year of publication; another similar paper on the electrolyte of the batteries (Croce et al. 1998) is in the top 0.1% of cited papers based on the same criteria. Because these fundamental papers pursue the advancement of science, many papers related to them by citation or topic are also highly cited.

However, in the field of rechargeable lithium batteries, there is also a huge amount of papers that report *incremental* innovations that address construction details of the batteries or their applications to different devices, for example, "*An implantable power supply with an optically rechargeable lithium battery*" (Goto et al. 2001). Papers of this type are cited in patents but normally receive a low number of citations in scientific journals. Therefore, if we could isolate this type of publication, their corresponding $e_p$ constant (and $P_{\text{top } 10\%}$/P ratio) would be low.

This reasoning is a conjecture that can be explained numerically. Consider a technical field with 100,000 publications per year and that there is a highly regarded annual *X Prize* that expert reviewers assign to the best paper; normally one in the top 0.005% of cited papers. Now, consider an advanced country that publishes 15,000 papers per year and that, on average, is awarded an *X Prize* every two years. However, because this country is highly focused on technology, 10,000 out of these 15,000 papers address incremental innovations in devices in the field. The constant $e_p$ when considering the total number of publications (15,000) is 0.05, so this advanced country looks like a developing country. Indeed, applying Eq. 2 (with $x = 0.005\%$), we calculate that, on average, this country will be awarded with one *X Prize* every 26 years. This prediction is far from reality, which implies that the bibliometric prediction fails. This failure occurs because this calculation considers that all the poorly cited papers report research that addresses advances of scientific knowledge that underpin the field in this imaginary example. However, this assumption is not true because 10,000 papers address technical details regarding the production of devices.

If we could to separate these two groups of papers, viz. the 10,000 technological papers with an $e_p$ constant of 0.015 and the 5,000 scientific papers with an $e_p$ constant of 0.12, we could perform this calculation correctly. Because the technological papers do not count toward the *X Prize*, they should not be included in the calculations with Eq. 2 regarding *fundamental* research; indeed, on the basis of these technological papers, the country would be awarded with an *X Prize* every 6,900 years. Meanwhile, based on the 5,000 scientific papers, the country would receive an *X Prize* every two years, which is the actual observation.

This conjecture distinguishing *fundamental* and *incremental* papers is actually a hypothesis that can be tested empirically. If it is correct, then there will be countries with high-technology industries and frequent Nobel laureates in which the standard bibliomertric indicators do not reveal this success (Section 7.3).

**4. Aim of this study**

Many studies have demonstrated that scientific wealth depends on economic wealth (Cimini et al. 2014; Allik et al. 2020; Cole and Phelan 1999; Docampo and Bessoule 2019; Gantman 2012; King 2004; May 1997; Mueller 2016; Rousseau and Rousseau 1998; Rahman and Fukui 2003), and in recent times, e.g. last 50−60 years, all the scientific achievements that have been awarded with Nobel Prizes have come from rich and highly developed countries, with Federico Leloir from Argentina being perhaps the only exception to this general rule. All this raises many questions. For example: Do only rich countries contribute to the progress of knowledge? What is the threshold economic level that allows countries to build a research system that contributes to the progress of knowledge? How important is research policy in the capacity of countries to contribute to the progress of knowledge?

Against this background, the aim of this study is to investigate the independent effects of economic wealth and all other factors, which we subsume into research policy, on the success of countries in creating new knowledge. In the first part of the study, we avoid fixing a single level of citation to measure such success but rather consider all possible levels, from the total number of papers to the top 0.01% of cited papers: P, $P_{top\ 10\%}$, $P_{top\ 1\%}$, $P_{top\ 0.1\%}$, and $P_{top\ 0.01\%}$. We also study the corresponding size-independent indicators $P_{top\ 10\%}/P$, $P_{top\ 1\%}/P$, $P_{top\ 0.1\%}/P$ and $P_{top\ 0.01\%}/P$.

In the final part of this study, we address the hypothesis described in the previous section by considering the number of triadic patent families as an indicator of industrial and technological activity. It is worth noting that we use this number exclusively as a reasonably indicator but do not establish any numerical relationship between the numbers of patents and technological publications. This will require further studies.

Overall, our aim is not to answer the questions posed above for specific countries, but rather to generate a general model based on a large number of countries. In some scatter plots we indicate specific countries, but this is only to document our reasoning or discuss the results. Analysis of specific countries lies beyond the scope of this study.

## 5. Methods

*5.1. Bibliometric data*

Our study takes advantage of the large amount of rigorous information provided by the Leiden Ranking for universities in terms of the citation-based distribution of papers in global top percentiles. Here, we use the number of papers in total and in a series of top percentile indicators, from the top 10% to the top 0.01%. This approach takes advantage of the suitability of top percentiles for research assessment (Bornmann et al. 2013) and the simplicity of their mathematical treatment because the results of all percentile counts are linked by a simple power-law function as described above (Section 2). It is worth noting that percentile evaluations are validated against peer review (Traag and Waltman 2019; Rodríguez-Navarro and Brito 2020a). To obtain the level of a country from the data in the Leiden Ranking, we aggregated the recorded P and $P_{top\ x\%}$ data of the universities belonging to that country (Supplementary Table S1).

The Leiden Ranking records full and fractional counting, but strong evidence supports that fractional counting describes research performance more accurately, not only at the address level in country and institution assessments (Rodríguez-Navarro 2012; Aksnes et al. 2012) but also at the author-level for individual assessments (Kolun and Hafner 2021). Waltman and van Eck (2015, p. 872) make a clear recommendation about fractional counting: "We therefore recommend the use of fractional counting in bibliometric studies that require field normalization, especially in studies at the level of countries and research organizations."

The downloaded data from the Leiden Ranking 2021 (https://www.leidenranking.com/; August 21, 2021) is an Excel file that contains the bibliometric data of 1,225 universities from 69 countries in six research fields over eleven four-year periods. For the purposes of this study, we selected the "Physical sciences and engineering" field and "fractional counting," extracting the data for the number of papers in the four top percentiles 1, 5, 10, and 50 ($P_{top\ 1\%}$, $P_{top\ 5\%}$, $P_{top\ 10\%}$, and $P_{top\ 50\%}$). Unless otherwise stated, we selected the first and the last periods recorded in the Leiden Ranking: 2006–2009 and 2016–2019. The Leiden Ranking does not include the $P_{top\ 0.1\%}$ and $P_{top\ 0.01\%}$ indicators, so we used Eq. 2 for their calculation, applying $P_{top\ 10\%}/P$ as a proxy for the constant $e_p$ (Eq. 3). In several countries $P_{top\ 1\%}$ had to be calculated because the values reported in the Leiden Ranking were very low; for homogeneity, we used this calculation for all countries (Rodríguez-Navarro and Brito 2021).

The Leiden Ranking includes universities from 69 countries that have produced at least 800 publications in the period 2016–2019, meaning that no institutions are included, and not all universities. However, our conclusions are extended to the whole country's research system. This extension is justified because research at universities represents a high proportion of countries' research at the highest level. For example, most Nobel laureates performed their awarded work at universities (Schlagberger et al. 2016), most highly cited researchers work in universities (Bornmann and Bauer 2015) and it is accepted that universities play a central role in the of knowledge production system (Godin and Gingras 2000).

However, considering the limitations of the database, we checked that our preference for the Leiden Ranking data did not introduce a bias into this study. For this purpose, we downloaded the country data from InCites (Clarivate Analytics) for the period 2016–2019, selecting the research areas that approximately make up the Leiden Ranking field of "Physical sciences and engineering." We then compared the countries' P and $P_{top\ 10\%}$ values (the two most important parameters in our study) between the two databases. For P, the Incite values were approximately five times higher but highly correlated with the Leiden Ranking values. Only three countries deviated from this general trend: China, Russia, and India. Eliminating these countries and the USA because it was an outlier, the Pearson correlation coefficient was 0.96 (two-sided p-value $< 10^{-10}$). For $P_{top\ 10\%}$ values, the Pearson correlation coefficient for all countries was 0.88 (two-sided p-value $< 10^{-10}$).

In conclusion, for the current purposes, our preference of using the Leiden Ranking versus the Incite data would not affect the essence of the results. Throughout this paper, we use the Leiden Ranking nomenclature for the indicators.

*5.2. Triadic patent families and other data*

As described in the previous section, in the analysis of *incremental* innovations, we considered the industrial and technological activity of countries. The number of triadic patent families per country was obtained from the OECD (doi: 10.1787/6a8d10f4-en; accessed 01 October 2021). For statistical reasons, we selected 31 countries with more than 10 patent families per year and eliminating the countries with high annual variability over the last five years with respect to a regression line fit.

The number of inhabitants, GDP (current US$) and GDP per capita were downloaded from the World Bank (https://data.worldbank.org/, August 23, 2021). The data for Taiwan, which are not recorded by the World Bank, were downloaded from https://countryeconomy.com/gdp/taiwan (August 25, 2021).

**6. Results**

*6.1. A few countries create a high proportion of the global new knowledge*

The most remarkable characteristic of the global scientific progress is nonuniformity across countries. For example, Schlagberger et al. (2016) identified 155 Nobel laureates from 1994 to 2014, but considering the affiliation when the Nobel Prize was awarded to them only seven countries accounted for three or more Nobel laureates; similarly, King (2004) found that only eight countries produce about 85% of the top 1% most cited publications. To study this inequality, we investigated which countries account for 90% of the global values of: P, $P_{top}$ $_{10\%}$, P$_{top\ 1\%}$, P$_{top\ 0.1\%}$ and P$_{top\ 0.01\%}$. The results (Table 1) support the notion of scientific inequality across countries, indicating that the number of countries (ordered from higher to lower values of each indicator) decreases with the decrease of the top percentile level.

Table 1. Countries that together account for 90% of the counts P, P$_{top\ 10\%}$, P$_{top\ 1\%}$, P$_{top\ 0.1\%}$ and P$_{top\ 0.01\%}$ with reference to the total values for the 69 countries under investigation

| 2006-2009 | | | | |
|---|---|---|---|---|
| P | P$_{top\ 10\%}$ | P$_{top\ 1\%}$ | P$_{top\ 0.1\%}$ | P$_{top\ 0.01\%}$ |
| USA | USA | USA | USA | USA |
| China | China | UK | UK | UK |
| Japan | UK | China | Germany | Germany |
| UK | Germany | Germany | China | Netherlands |
| Germany | Japan | France | France | Switzerland |
| South Korea | France | Japan | Netherlands | France |
| France | Canada | Canada | Switzerland | China |
| Italy | Spain | Netherlands | Canada | Canada |
| Spain | Italy | Switzerland | Australia | Australia |
| Canada | South Korea | Spain | Spain | Spain |
| India | Australia | Italy | Japan | Denmark |
| Taiwan | Netherlands | Australia | Italy | |
| Australia | India | South Korea | Denmark | |
| Brazil | Switzerland | Sweden | | |
| Poland | Taiwan | India | | |
| Iran | Sweden | Singapore | | |
| Turkey | Turkey | Denmark | | |
| Netherlands | Singapore | | | |
| Sweden | Iran | | | |
| Switzerland | Brazil | | | |
| Israel | | | | |
| Singapore | | | | |
| 2016-2019 | | | | |
| China | China | China | USA | USA |
| USA | USA | USA | China | China |
| Germany | UK | UK | UK | UK |
| UK | Germany | Germany | Australia | Australia |
| Japan | South Korea | Australia | Germany | Singapore |
| South Korea | Australia | France | Singapore | Switzerland |
| India | Japan | Canada | Switzerland | Germany |
| Iran | France | Singapore | Netherlands | Netherlands |
| France | Iran | Switzerland | France | France |
| Italy | India | South Korea | Canada | Canada |
| Canada | Canada | Italy | Italy | Italy |
| Spain | Italy | Netherlands | South Korea | |
| Australia | Spain | Iran | Iran | |
| Brazil | Switzerland | Japan | Spain | |
| Poland | Netherlands | India | | |
| Taiwan | Singapore | Spain | | |
| Turkey | Brazil | | | |
| Netherlands | Sweden | | | |
| Sweden | Poland | | | |
| Switzerland | | | | |
| Russia | | | | |
| Singapore | | | | |

For the period 2006–2009, only 22 out of the 69 countries account for 90% of all published papers, whereas the number decreases to 11 countries accounting for 90% of the global papers in the top 0.01% of cited papers. Furthermore, with the decrease of the top percentile level some countries disappear while others appear in the lists. For example, Denmark is not in the list by P (lying in position 29, Table S2) but appears at position 11 in the $P_{top\ 0.01\%}$ list. In contrast, India and Brazil are in positions 12 and 15 in the list by P, respectively, but are not in the list by $P_{top\ 0.01\%}$ (in positions 20 and 30 in Table S2). Similar conclusions can be drawn from the data for the period 2016–2019.

A comparison of the two periods reveals a notable improvement of research in two countries: China and Singapore, both of which are in the lists of countries accounting for 90% of the five indicators (P and the four top percentiles). In the period 2006–2009, a comparison of the position of these countries with respect to the UK, Germany, Switzerland, and other countries shows that they continuously drop in position with decreasing percentile value. In contrast, in a similar comparison for the period 2016–2019, China only switches its first position with the USA, while Singapore rises from position 22 by P to position 5 by $P_{top\ 0.01\%}$, overtaking Germany and France.

Table 2. Percent of the number of inhabitants or GDP of the countries that account for 90% of the counts of P, $P_{top\ 10\%}$, $P_{top\ 1\%}$, $P_{top\ 0.1\%}$ and $P_{top\ 0.01\%}$ (Table 1) with reference to the corresponding total values for the 69 countries under investigation

|  | Inhabitants | | GDP | |
| --- | --- | --- | --- | --- |
|  | 2006-2009 | 2016-2019 | 2006-2009 | 2016-2019 |
| P | 75 | 76 | 84 | 86 |
| $P_{top\ 10\%}$ | 74 | 71 | 83 | 82 |
| $P_{top\ 1\%}$ | 67 | 68 | 79 | 78 |
| $P_{top\ 0.1\%}$ | 43 | 41 | 73 | 68 |
| $P_{top\ 0.01\%}$ | 39 | 37 | 60 | 63 |

Next, we calculated the relationship between the cumulative values of population and GDP of the countries accounting for 90% of the total values of the investigated indicators. Table 2 presents the results, which are similar for the two periods. Regarding the values of P, the countries accounting for 90% of its total value contribute 75% of the global population and 85% of the global GDP, suggesting that population size or GDP reasonably explains the number of publications. For the other indicators ($P_{top\ 10\%}$, $P_{top\ 1\%}$, $P_{top\ 0.1\%}$ and $P_{top\ 0.01\%}$), the percent of population corresponding to the most productive countries decreases with the decrease of the top percentile value, approximately in parallel to the decrease of the number of countries. In contrast, the decrease of the percent of the total GDP corresponding to the countries accounting for 90% of the indicator is much lower. For example, in both periods, there are 22 countries in the list by P and 11 in the list by $P_{top\ 0.01\%}$ from the list of 69 countries; the corresponding population of the countries that account for 90% of the total indicator is 75% for P and 38% for $P_{top\ 0.01\%}$ (half the number of countries and half the number of inhabitants). In contrast, the corresponding GDP of the countries accounting for 90% of the global indicator is 85% for P and 62% for $P_{top\ 0.01\%}$ (half the number of countries but only 27% lower in GDP).

In summary, the wealthiest countries create most of new knowledge.

*6.2. Economic and scientific wealth*

*6.2.1. Size-dependent analyses*

Next, we investigated the dependence of P, $P_{top\ 10\%}$, $P_{top\ 1\%}$, $P_{top\ 0.1\%}$, and $P_{top\ 0.01\%}$ on GDP across countries; Figures 1 and 2 show plots of P, $P_{top\ 1\%}$, and $P_{top\ 0.01\%}$ versus the GDP of the countries for the periods 2006–2009 and 2016–2019, respectively, excluding USA from Figure 1, and USA and China from Figure 2 because of the outlier positions of their GDP values.

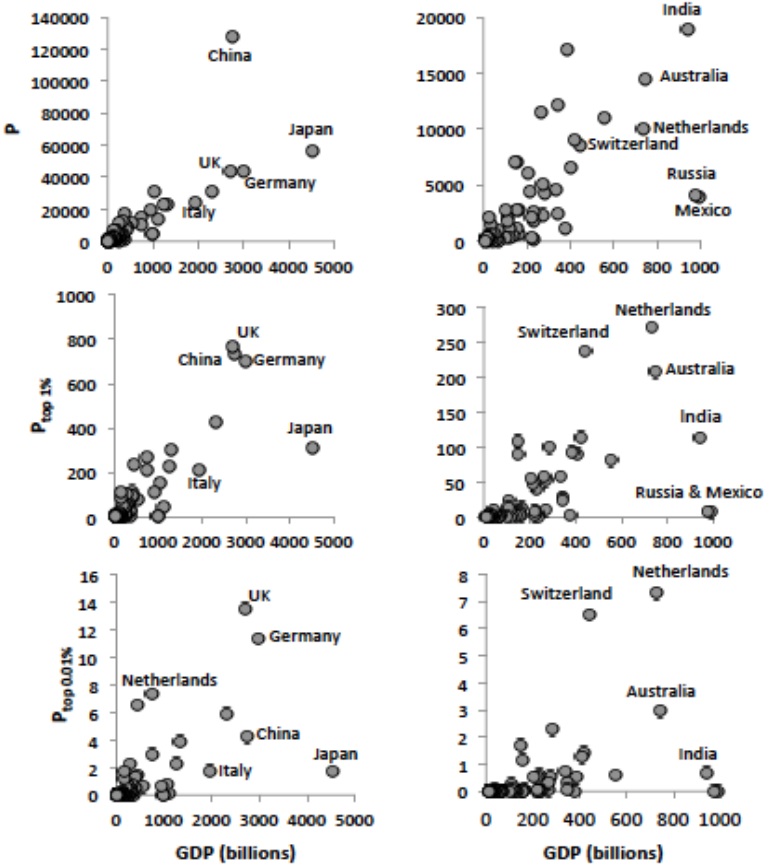

Figure 1. Scatter plots of P, $P_{top\ 1\%}$, and $P_{top\ 0.01\%}$ versus GDP for the period 2006–2009. The USA is omitted owing to its outlier position. The plots on the right correspond to countries whose GDP is less than 1,000 US$.

Visual inspection of these plots reveals a clear dependence of P on GDP, albeit with notable dispersion of the data points; this dispersion increases for smaller top percentiles and is very high for $P_{top\ 0.01\%}$. The Pearson and Spearman rank correlation coefficients confirm the existence of a correlation for both periods (2006–2009 and 2016–2019). For example for 2006–2009: for P, the Pearson correlation coefficient excluding the USA is 0.81 (two-sided p-value $< 10^{-10}$) and the Spearman correlation coefficient of all the data is 0.87 (two-sided p-value $< 10^{-10}$); and for $P_{top\ 0.01\%}$, the Parson correlation coefficient excluding the USA and China is 0.64 (two-sided p-value $= 3.3 \cdot 10^{-9}$) and the Spearman correlation coefficient of all the data is 0.80 (two-sided p-value $< 10^{-10}$).

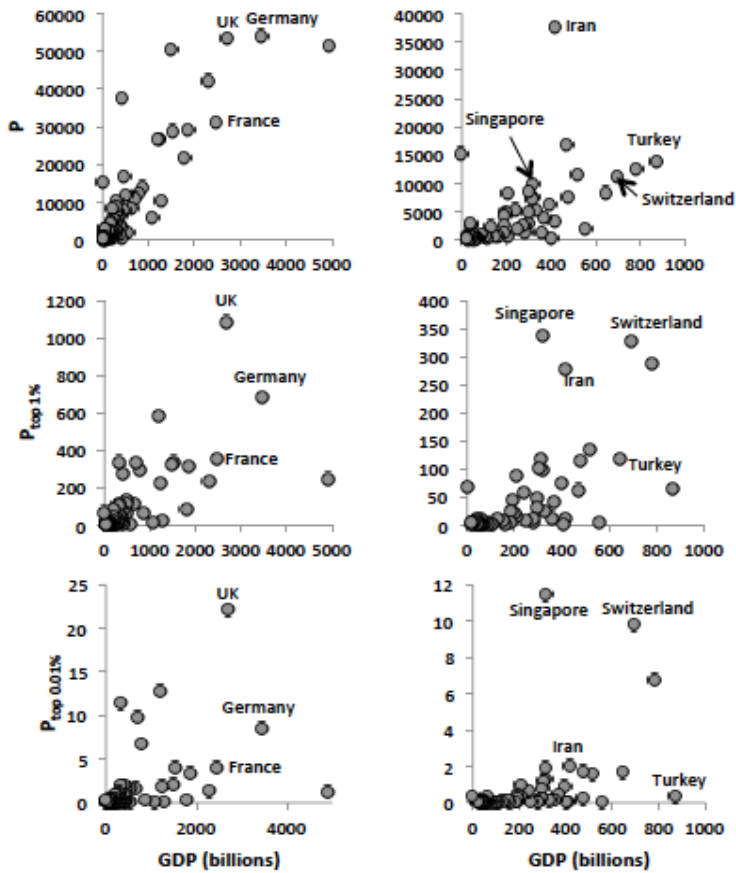

Figure 2. Scatter plots of P, $P_{top\ 1\%}$ and $P_{top\ 0.01\%}$ versus GDP in the period 2016–2019. The USA and China are omitted owing to their outlier positions. The plots on the right correspond to countries whose GDP is less than 1,000 US$

These results demonstrate that economic wealth is an important determinant of scientific wealth. However, despite these correlation coefficients, it is notable that the order of the countries' data points is different in each case: i.e. the relative positions of countries with similar GDPs change depending on the stringency of the $P_{top\ x\%}$ indicator considered. For example, in Figure 1, the relative positions of India, Australia and the Netherlands, or the position of Switzerland with reference to other countries, and in Figure 2, the change of the relative positions of Switzerland and Turkey, or Singapore and Iran.

To analyze the strong correlations described above, one must consider that the range of variation of the data across the 69 countries of this study is very high: three orders of magnitude in GDP and five orders of magnitude in $P_{top\ 0.01\%}$; these large variations and the large number of countries might conceal important information. As shown above (Section 6.1) the number of countries that make significant contributions to the global new knowledge acquisition is low, which implies that studying these countries is likely to provide more accurate information than studying all countries.

To further investigate this issue, we ordered the countries by their number of publications. For the period 2006–2009, we ordered the countries by their total number of papers (P) and selected the 22 countries publishing from 56,000 to 6,000 papers (a range of one order of magnitude). For these countries, the correlation between P and GDP is very high (Figure 3A; Pearson correlation coefficient of 0.94, two-sided p-value $1.5 \cdot 10^{-8}$); the same occurs if the countries are ordered by $P_{top\ 10\%}$ (results not shown). In contrast, if we order the countries by their $P_{top\ 0.01\%}$ and select the first 16 countries (for which the indicator varies by one order of magnitude, from 13.5 to 1.2), $P_{top\ 0.01\%}$ and GDP do not show a correlation (Figure 3B; Pearson correlation coefficient 0.38, two-sided p-value 0.14). Similar results are obtained for the period 2016–2019.

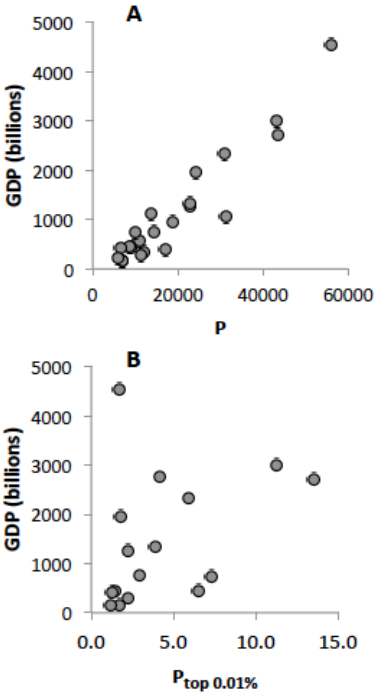

Figure 3. Scatter plots of the GDP versus P and $P_{top\ 0.01\%}$ of top countries ordered by these parameters (22 in A and 19 in B) for the period 2006–2009.

In summary, GDP is a strong determinant of the value of all indicators if all the countries are considered. It is also a strong determinant if a small set of countries is selected by considering the highest values of P and $P_{top\ 10\%}$. In contrast, if the set of countries is selected based on to the highest $P_{top\ 0.01\%}$ values, the value of this indicator is not correlated with the GDP.

*6.2.2. Size-independent analyses*

The findings described above (Figure 3B) raise the possibility that the determinant of research success at the *fundamental* level ($P_{top\ 0.1\%}$ and $P_{top\ 0.01\%}$) is economic measured not in terms of GDP, but in terms of GDP per capita. The rationale is that, across countries, only those with high GDP per capita will have the capacity to invest in research and in the other functional structures necessary for successful research. We therefore next investigated this possibility. However, GDP per capita is size independent and must be compared with size-independent research indicators.

To maintain consistency with the analyses above, the size-independent indicators $P_{top\ 10\%}/P$, $P_{top\ 1\%}/P$, $P_{top\ 0.1\%}/P$ and $P_{top\ 0.01\%}/P$ were used. However, according to Eq. 2 and considering that $P_{top\ 10\%}/P$ is a proxy of $e_p$, these size-independent indicators are proxies of $e_p$, $e_p^2$, $e_p^3$ and $e_p^4$. This implies that, among countries in the scatter plots of these indicators versus GDP per capita, the relative positions of countries will always be the same while only the distance between them will vary (Rodríguez-Navarro and Brito 2021). This also implies that all the correlation coefficients between GDP per capita and the indicators will be equal. Figure 4 shows the scatter plots of countries based on their $P_{top\ 10\%}/P$ ratio and GDP per capita for the periods 2006–2009 and 2016–2019. Visual inspection of the scatter plots for the total number of countries (left panels) indicates that the values of the indicator show a clear dependence on GDP per capita; the Pearson correlation coefficients are 0.77 (two-sided p-value < $10^{-10}$) for 2006–2009 and 0.83 (two-sided p-value < $10^{-10}$) for 2016–2019.

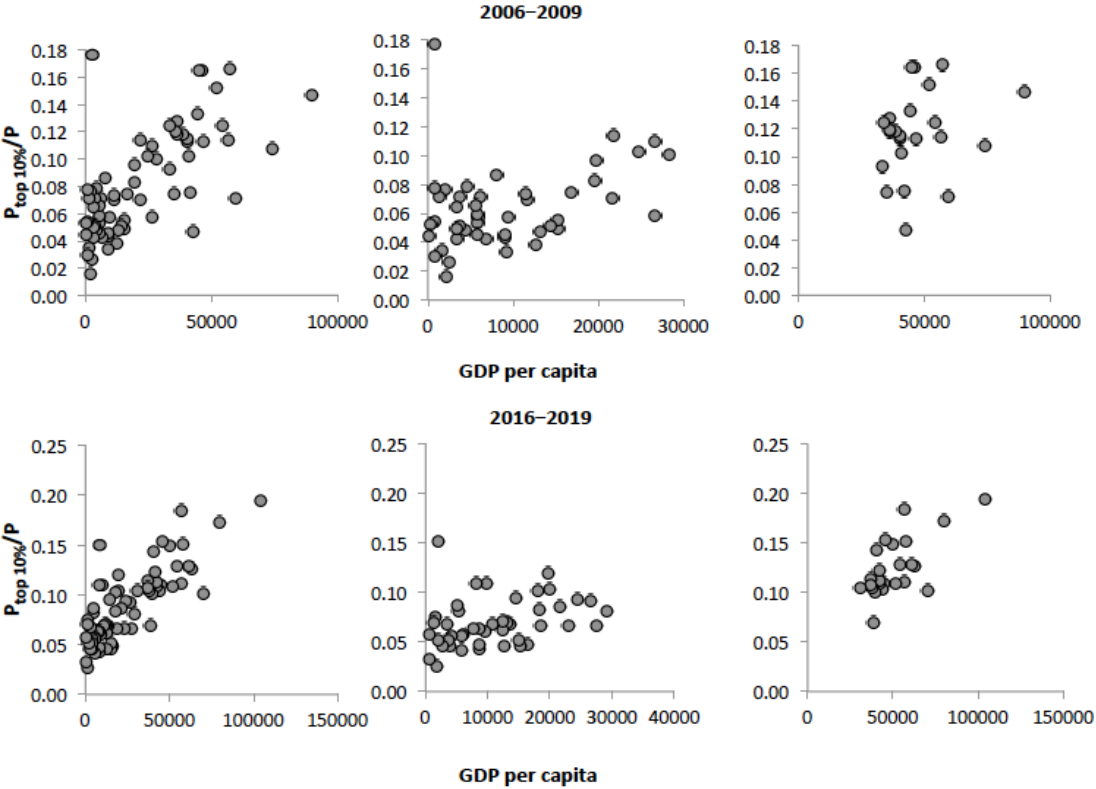

Figure 4. Scatter plot of countries according to their GDP per capita and $P_{top\ 10\%}/P$ ratio. Left panels, all countries; two right panels, countries divided according to the GDP per capita: up to 30,000 and above 30,000 US$

However, following a similar reasoning as in the previous section regarding the wide range of data when considering all countries, we divided the lists of countries into two groups, having GDP per capita above and below 30,000 US$. Visual inspection of the scatter plots indicates that a correlation does not exist or that it is very poor in either group (Figure 4).

In summary, considering all countries, the size-independent parameters $P_{top\ x\%}/P$ and GDP per capita are highly correlated, but taking sets of countries in narrow ranges of GDP per capita, the correlations are very weak or inexistent.

*6.3. Patenting activity*

To distinguish between countries with high versus low technological activity, we used the number of triadic patent families as a convenient indicator; we selected 31 countries with more than 10 patent families in 2018 that also showed good stability of the indicator (Section 5.2). For analytical purposes, we divided these countries into two sets based on the ratio of the number of patents to GDP per capita (Table S3): a first set of 13 countries with 3.8 to 0.8 patents per billion US$ of GDP and a second set of 18 countries with 0.6 to 0.1. We also included the number of papers in this analysis. Japan and the USA were outliers based on the number of patent families, and the USA and China based on the number of papers.

As a first approach, we tested whether the previously observed correlation between GDP and the number of papers (Figure 1 and 2) was retained in these two sets of countries (comprising 12 and 16 countries after omitting outliers). The finding is that the two sets of countries are mixed and all countries follow the same trend (Figure 5A). In contrast, in the scatter plot of the number of triadic patents versus the number of papers (Figure 5B), the two sets of countries behave differently. For both sets, the number of patents increases with increasing number of papers, but this increase is faster in the set with higher patenting activity with respect to GDP.

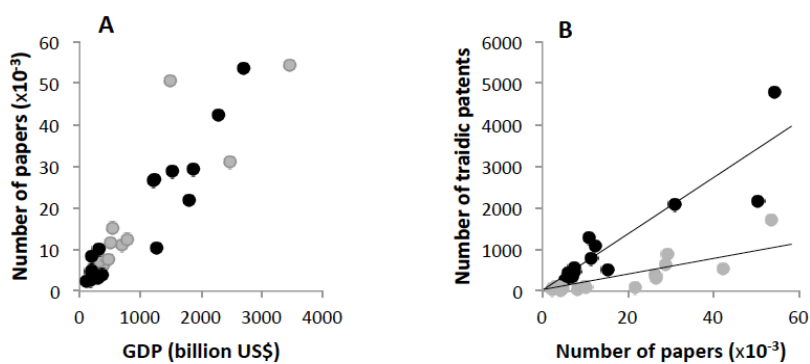

Figure 5. GDP dependence of the number of papers (A) and relationship between the numbers of papers and triadic patent families (B). Black circles, high patenting countries; grey circles, low patenting countries. Period of evaluation, 2016–2019, GDP and patents in 2018. The USA and China are exclude because as outliers. Lines in B are drawn as a guide to the eye

To further investigate the position of each country regarding these two products of its research (publishing and patenting) we used two size-independent indicators: the $P_{top\ 10\%}/P$ ratio (publishing efficiency) and the number of patents per billion US$ of GDP (patenting efficiency). Figure 6 shows the distribution map of the 31 countries investigated. Japan is way ahead of other countries in patenting activity, while the other countries are distributed across the whole surface of the map, albeit with a notable proportion in the lower left part, with 11 countries showing patenting activity ten times lower than that of Japan and a $P_{top\ 10\%}/P$ ratio slightly above or below 0.1.

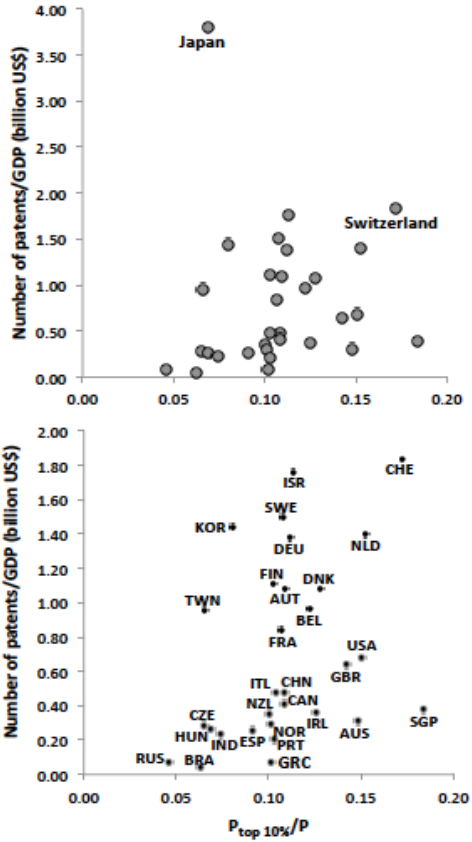

Figure 6. Scatter plot of countries according to their P_top 10%/P ratio and number of triadic patent families per billion US$ of GDP

If our hypothesis about the lower $P_{top\ 10\%}/P$ ratio in countries with high technological activity is correct (Section 3), countries with high technological activity might show a lower $P_{top\ 10\%}/P$ ratio than actually corresponds to their research efficiency. This may be the case of Japan, South Korea, and Taiwan.

*6.4. Research efficiency in low patenting countries*

In the absence of a convenient numerical correction of the $P_{top\ 10\%}/P$ ratio to account for technological activity (Section 4), analysis of research efficiency and its relationship with GDP per capita remains difficult. We therefore focus our attention on the 11 countries in the lower left part of the country map in Figure 6. In these countries, patenting is low and it is improbable that technological activity will have a significant effect on the $P_{top\ 10\%}/P$ ratio. Considering GDP per capita, the presence in this group of Norway and even Spain is unexpected, so we investigated in more detail the research of these two countries in comparison with Portugal and Greece, including Singapore as a control country (the GDP per capita in 2018 of these five countries being 70,459, 26,505, 19,978, 18,117 and 56,828 US$, respectively).

Figure 7A shows the evolution of P and Figure 7B the evolution of the $P_{top\ 10\%}/P$ ratio through the 11 periods recorded in the Leiden Ranking. Except for Greece (where P decreased), P increased over time in the other four countries. In contrast, the evolution of the $P_{top\ 10\%}/P$ ratio is more complex. In Greece, the $P_{top\ 10\%}/P$ ratio remains quite stable despite the decrease of P; in Norway and Portugal, the $P_{top\ 10\%}/P$ ratio show oscillations but remains stable overall, despite the growth of P; in Spain, the $P_{top\ 10\%}/P$ ratio remains stable over the six periods but then decreases, again despite the growth of P. Finally, in Singapore, the $P_{top\ 10\%}/P$ ratio increases in parallel with the an increase of P.

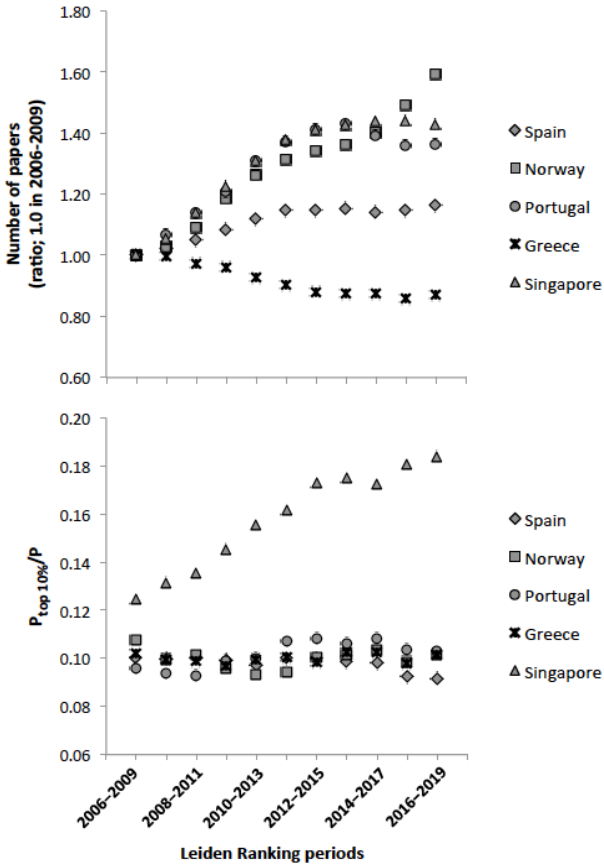

Figure 7. Temporal evolution of the number of papers and the $P_{top\ 10\%}/P$ ratio in selected countries

In summary, it seems that, independently of economic wealth, the research policy of some countries is aimed at increasing the number of papers and not to increase the $P_{top\ 10\%}/P$ ratio.

## 7. Discussion

A large number of studies have established that the scientific wealth of countries depends on their economic wealth (e.g. Cimini et al. 2014; Allik et al. 2020; Cole and Phelan 1999; Docampo and Bessoule 2019; Gantman 2012; King 2004; May 1997; Mueller 2016; Rousseau and Rousseau 1998; Rahman and Fukui 2003; Satish 2021). However, it can be intuitively expected that factors other than economic ones could also determine the capacity of countries to contribute to the global acquisition of knew knowledge. As explained in Section 4, the aim of the current study is to investigate whether other factors, independent of economic ones, determine the research efficacy of countries. We subsume all these factors into the broad term of research policy, but some of the noneconomic factors conditioning the research performance of countries may not be related to their research policy (e.g. the stock of knowledge). For example, Khosrowjerdi and Bornmann (2021) investigated national culture and found a correlation between survival versus self-expression values and $PP_{top\ 10\%}$ (equal to $100 \cdot P_{top\ 10\%}/P$). However, it seems unlikely that anything other than research policy could explain why, over the years, some countries have simultaneously increased their number of papers and the $P_{top\ 10\%}/P$ ratio while others have increased only the number of papers (Figure 7).

*7.1. Indicators of new knowledge creation*

So far, there is no agreement regarding which indicator should be used to assess new knowledge creation. The number of papers, number of citations, and number of papers in the top 10% or 1% of cited papers are the indicators related to scientific wealth most commonly applied in the above-cited studies. Among these studies, two similar to the present one, published in the same year (2018), one used the top 1% (Vinkler 2018) while the other the top 10% (Sandström and van den Besselaar 2018) of cited papers as a reference standard.

As suggested above, this uncertainty regarding the most appropriate bibliometric indicator to use in studies of research assessment is accompanied by a lack of consensus regarding the type of new knowledge to be assessed: either *fundamental* or *incremental* (Section 2). Although there is no fixed top percentile that separates these two types of new knowledge, researchers in many fields would probably agree that less than 1 paper out of 1,000 is truly *fundamental*.

This observation leads to the conclusion that a complete appraisal of the research success of a country cannot be achieved using a single indicator, be it based on the number of either citations or papers. In the case of top percentiles, a highly stringent top percentile (e.g. the top 0.01% or 0.1%) will capture the acquisition of *fundamental* new knowledge but will ignore thousands of papers and will not capture the acquisition of *incremental* new knowledge. In contrast, a less stringent top percentile (e.g. the top 10%) will capture the acquisition of *incremental* new knowledge, but the high number of papers at this level will conceal the information needed to assess the acquisition of *fundamental* new knowledge.

Although the assessment of only *fundamental* new knowledge can be informative in some cases (e.g. Rodríguez-Navarro 2011), to achieve a full assessment of research, two different indicators seem necessary: either P or $P_{top\ 10\%}$ as a reasonably indicator for the assessment of *incremental* new knowledge, and either $P_{top\ 0.1\%}$ or $P_{top\ 0.01\%}$ for the assessment of *fundamental* new knowledge. A single indicator, e.g. $P_{top\ 1\%}$, cannot be comprehensive because, if two countries have different P, the same $P_{top\ 1\%}$ may correspond to two different proportions of *incremental* (P or $P_{top\ 10\%}$) and *fundamental* ($P_{top\ 0.1\%}$ or $P_{top\ 0.01\%}$) new knowledge.

Applying Eq. 2 to this reasoning, the two indicators that define the activity of a country in new knowledge creation are P and $e_p$. The former describes the amount of research, and the latter the efficiency of the country in publishing papers that are highly cited (Rodríguez-Navarro and Brito 2018).

In summary, although the indicators for *incremental* and *fundamental* new knowledge are P or $P_{top\ 10\%}$ and $P_{top\ 0.1\%}$ or $P_{top\ 0.01\%}$, respectively, the research performance of a country is fully described by P and $P_{top\ 10\%}/P$.

*7.2. Scientific and economic wealth*

Consistent with previous studies cited above, the results of the current study (Figures 1, 2 and 5A) reveal that the number of papers (P) and $P_{top\ 10\%}$ correlate with the GDP of countries, indicating that the research activity of countries is a function of their economic size. Although some variability exists, it is not high and remains similar in countries with high or low technological activity (Figure 5A). In summary, GDP provides reasonable information about the total amount of new knowledge that a country normally creates.

However, because a single bibliometric indicator is insufficient to define the overall research performance of a country (as shown in the previous section), a single economic parameter also does not suffice. The second bibliometric indicator that defines the research performance of a country is the $P_{top\ 10\%}/P$ ratio, and it is shown in Section 6.2.2 that, in a broad sense, GDP per capita is an important determinant of the $P_{top\ 10\%}/P$ ratio. However, while GDP provides reasonable information about the amount of new knowledge that a country creates, GDP per capita is a poor determinant of research efficiency ($P_{top\ 10\%}/P$ ratio; Figure 4). For example, for the period 2016–2019, two countries with GDP per capita below 10,000 US$ (China and Malaysia) achieved the same $P_{top\ 10\%}/P$ ratio (0.11) as France, Sweden, Canada, Austria, Germany, and Israel, and higher values than Italy, Norway, and New Zealand. Among all these countries, GDP per capita varies from 30,000 to 70,000 US$ (Table S4).

These observations imply the existence of important factors that modify the economic dependence of the $P_{top\ 10\%}/P$ ratio.

*7.3. Incremental knowledge can serve either scientific or technological progress*

Section 3 describes how *incremental* new knowledge can play two roles, toward either scientific or technological advances, and that the citation functions corresponding to papers that support each function are different. This implies that research assessments of industrialized countries should consider both functions. To consider the technological activity of countries, we used their patenting activity as estimated by the number of triadic patent families.

Although *incremental* new knowledge plays the two roles described above, the analysis of the results shown in Figure 5A reveals that the number of papers published with respect to GDP is independent from the patenting activity. This finding implies that, if two countries have the same GDP but very different technological activity, they will create similar amounts of *incremental* new knowledge but that the country with higher technological activity will create less *incremental* new knowledge that is addressed towards *fundamental* new knowledge than the country with lower technological activity. To put this in figures, if two countries with GDP of 2,000 billion US$ publish 25,000 papers (Figure 5A) and one has 1,500 while the other has 100 triadic patent families, the number of papers addressed toward producing *incremental* innovations could be much higher in the former than the latter. From the opposite viewpoint, the total amount of knowledge created that is addressed toward the progress of *fundamental* knowledge could be much higher in the latter than in the former.

Under these circumstances, $P_{top\ 10\%}$ but not P will be affected by technological activity. Therefore, it can be conjectured that, if the ratio between the number of triadic patents and GDP is high, a low $P_{top\ 10\%}/P$ ratio in technologically advanced countries could conceal excellent scientific research (Section 3).

This hypothesis can be tested based on the country map shown in Figure 6, where the most notable case is Japan. It has the highest ratio of triadic patents to GDP in the world but its poor bibliometrics indicators fail to predict its high scientific level:

> National science indicators for Japan present us with a puzzlement. How can it be that an advanced nation, a member of the G7, with high investment in R&D, a total of 18 Nobel Prize recipients since 2000, and an outstanding educational and university system looks more like a developing country than a developed one by these measures? The citation gap between Japan and its G7 partners is enormous and unchanging over decades. Japan's underperformance in citation impact compared to peers seems unlikely to reflect a less competitive or inferior research system to the degree represented (Pendlebury 2020, p. 134).

Japan therefore confirms our hypothesis, which is also confirmed by two other cases: Germany and France. According to their position in Figure 6 ($P_{top\ 10\%}/P$ around 0.11), it should be almost impossible for these countries to be repeatedly awarded Noble Prizes in natural sciences, as is the case. According to the number of Nobel Prizes in natural sciences, Germany is slightly below the UK, and France is only slightly lower than Germany. This again is coincident with the prediction of our hypothesis, considering that the number of patents is higher in Germany and France than in the UK. In two other countries (Sweden and Israel), the number of Nobel laureates is low because they are small countries, but their low $P_{top\ 10\%}/P$ ratios around 0.11 are incompatible with their position in terms of the number of Nobel laureates.

In summary, the hypothesis that technological activity could mislead the bibliometric assessment of the scientific wealth of countries is consistent with the empirical data.

Therefore, analysis of the country map shown in Figure 6 should be performed by considering the $P_{top\ 10\%}/P$ ratio and patenting activity simultaneously. However, our data cannot distinguish between two hypothetical types of countries: those that publish a significant amount of their patented technological advances, and others that do so very seldomly.

Even with these caveats, the map in Figure 6 supports an interesting analysis: Singapore shows a very high $P_{top\ 10\%}/P$ ratio but low patenting activity, suggesting that its research strategy is focused on the advancement of knowledge and revolutionary innovations and that the $P_{top\ 10\%}/P$ ratio describes its real scientific performance. Australia follows the same strategy, albeit with lower success than Singapore. The USA and the UK show intermediate patenting activity but high $P_{top\ 10\%}/P$ ratios, suggesting that, in addition to patenting, these countries are simultaneously focused on the advancement of knowledge and revolutionary innovations. Consequently, in comparison with Singapore, their real scientific level might be higher than suggested by their $P_{top\ 10\%}/P$ ratios. On the opposite side of the map, South Korea and Taiwan show high patenting activity but low $P_{top\ 10\%}/P$ ratios that conceal a probably high scientific level. Two countries (Switzerland and the Netherlands) show high $P_{top\ 10\%}/P$ ratios and patenting activity, which implies outstanding scientific levels or low publishing activity of technological advances.

In the lower left part of the map, countries have low patenting activity and low $P_{top\ 10\%}/P$ ratio, which might reveal a real low research performance. This applies to at least 10 countries: Russia, Brazil, the Czech Republic, Hungary, India, Spain, New Zealand, Norway, Portugal, and Greece.

Further research is needed to achieve a numerical correction of the $P_{top\ 10\%}/P$ ratio in countries with high technological activity and a culture of publishing incremental innovations in scientific journals. Applying such a correction would increase the accuracy of scientometrics and may address its criticisms (Marginson 2021)

*7.4. Research policy also counts*

In the discussion above about the research map of countries (Figure 6), we considered GDP but not GDP per capita or the dependence of the $P_{top\ 10\%}/P$ ratio from GDP per capita; this consideration introduce other features of research in each country. For example, the position of Brazil among the countries in the lower left of Figure 6 can be explained by its low GDP per capita, but the same cannot be said for Norway or even Spain. In these cases, neither high patenting activity nor low GDP per capita can explain the low $P_{top\ 10\%}/P$ ratio, and these two cases are not isolated. Indeed, by the same reasoning, many differences in the $P_{top\ 10\%}/P$ ratios between pairs of countries cannot be explained, e.g. Germany and the Netherlands. Furthermore, the countries shown in Figure 6 were selected because their numbers of triadic patent families are statistically robust. The $P_{top\ 10\%}/P$ ratios of the omitted countries with respect to those in Figure 5 exhibit high variability that cannot be explained by their patenting activity. All these cases suggest that, for similar GDP per capita, a factor other than technological activity acts on the $P_{top\ 10\%}/P$ ratio of countries, i.e. on the creation of *fundamental* new knowledge ($P_{top\ 0.01\%}$ or $P_{top\ 0.1\%}$). Currently, it is difficult to conceive that this factor is anything but research policy, as mentioned at the beginning of Section 7.

All these observations raise an interesting question about the effects of research policy on a country's research performance at the *fundamental* level ($P_{top\ 0.01\%}$ or $P_{top\ 0.1\%}$), because this level is not normally considered in studies on the links between economic and scientific wealth.

More closely related to our question regarding the effects of research policy on countries' research, Linda Butler, almost 20 years ago, tried to study the consequences of the Australian research policy on the quality of research outputs: "the academic response to the linking of funds, at least in part, to productivity measures undifferentiated by any measure of quality — publication numbers jumped dramatically, with the highest percentage increase in the lower impact journals" (Butler 2004, p. 389). This finding was interpreted as a decline of Australian research, but more recently van den Besselaar et al. (2017) revisited the Australian case, showing that Butler's finding was incorrect. This study, and another study by Schneider (Schneider et al. 2016) calculated the $P_{top\ 10\%}/P$ ratios and showed that the Australian ratio has increased monotonically since the mid-1990s, consistent with the high Australian $P_{top\ 10\%}/P$ ratio found in Australian herein (Figure 6).

In relation to the current question, the Norwegian model of research funding has also been described (Sivertsen 2018) and studied, with contradictory interpretations of its effects on research results (Schneider et al. 2016; van den Besselaar and Sandström 2017). In (Schneider et al. 2016, Figure 7), the $P_{top\ 10\%}/P$ ratio for Norway increased from the mid-1990s until approximately 2009, after which it seems to have remained steady at around 0.11. We found approximately the same ratio, and the high contrast between the low Norwegian $P_{top\ 10\%}/P$ ratio but high GDP per capita for Norway suggests that Norwegian research might have an efficiency problem. Further supporting this possibility, Figure 7 shows that the evolution of P and the $P_{top\ 10\%}/P$ ratio for Norway during the 11 evaluation periods, revealing that P increases notably while the $P_{top\ 10\%}/P$ ratio remains almost constant, oscillating around 0.1. Because the GDP increased very little during the period 2006–2016, it seems that research policy in Norway is aimed at only increasing the number of papers but not the efficiency of research. Indeed, Portuguese and Norwegian research show similar evolutions even though the GDP per capita of Portugal is at least 3.5 times lower than that of Norway.

*7.5. Wrong research policies lead to scientific stagnation, the case of Spain*

As considered above, it seems hard to believe that the differences between countries described herein can be explained by anything other than research policy, including as research policy the scientific culture discussed by Godin and Gingras (2000). Probably many factors that differ among countries affect the researchers' attitude or country's research environment. Regarding the latter, it is worth noting that research projects that are generously funded by the European Research Council have a greater probability of success if they are executed in the UK, the Netherlands, or Switzerland than if executed in Germany, France, Spain or Italy (Rodríguez-Navarro and Brito 2020b).

Regarding the factors that can affect the research success of countries, Bornmann and Marx (2012) propose the "Anna Karenina Principle" that only a few factors differing in each case can result in the difference between countries. According to this, the causes of research failure are better studied case by case; the case of Spain is illustrative because, according to the Anna Karenina principle, its incorrect research policy (Rodríguez-Navarro 2009) is what has led to its failure to succeed.

In Spain, a successful research policy was initiated in the late 1980s, resulting in a notable growth in the number of publications, starting in 1990 (Jiménez-Contreras et al. 2003), much faster than the growth of its GDP. The specific component of the research policy that resulted in this growth is still under discussion (Osuna et al. 2011), but this is not relevant to the purpose of this study.

A change in Spanish research policy took place in the early 2000s, but in the wrong direction, using the impact factor of journals as a factor determining the success of researcher evaluations (Jiménez-Contreras et al. 2002). As a consequence, in 2009, it was evident that Spanish research was characterized by sound research that produced unimportant discoveries (Rodríguez-Navarro 2009). Since then, the whole research policy in Spain has been characterized by the extensive use of journal impact factors and the position of the journal in which papers are published in lists ordered by impact factor (Delgado-López-Cózar et al. 2021). Publication in journals in the first quarter of this list (Q1) is usually required for successful evaluations of individual researchers, although it is known that this approach is not rational for selecting the most influential papers (Brito and Rodríguez-Navarro 2019).

The second unfortunate research policy in Spain was to decrease the proportion of research projects that were funded in each call, even though the selection process does not have sufficient accuracy to establish a reasonable rejection threshold. As a consequence, many solid projects are rejected. Although this policy was made public in 2012 being attributed to the economic crisis, it was actually established several years before, always based on the argument of targeting improvement: "With will, our slimed-down R&D system will be able to take advantage of the crisis—and emerge from it stronger than ever" (Vela 2011). While such a policy of providing more funding to better evaluated projects is reasonable, that of not funding projects below an arbitrary threshold based on review metrics is damaging because such all-or-nothing funding based on an arbitrary threshold will leave many reasonably projects unfunded, especially when the review process is based on inappropriate indicators.

These two policy measures generate risk aversion among researchers (Zoller et al. 2014), who will only try to publish in high-impact journals, selecting low risk goals for their projects that will finally lead to poorly cited papers, or in other words, with a very low probability of creating *fundamental* new knowledge. The current results confirm this prediction; Figure 7 shows that, under this policy, the number of papers has grown while the $P_{top\ 10\%}/P$ ratio first remained stable at around 0.1 but then decreased during the last five years.

## 8. Conclusions

This study confirms a strong link between economic and research wealth. GDP conditions the amount of research in terms of the number of publications, while GDP per capita conditions the efficiency in terms of the $P_{top\ 10\%}/P$ ratio. Although in both cases there are deviations for some countries, these are much more substantial in the case of the $P_{top\ 10\%}/P$ ratio. Two country factors have strong effects on this ratio: the technological activity, which we estimate using the number of triadic patent families, and the research policy. The best example of the effect of technological activity is Japan, for which the $P_{top\ 10\%}/P$ ratio remains at the level of low-income countries despite its high scientific success if estimated based on the number of Nobel laureates. GDP per capita seems to impose a limit on the $P_{top\ 10\%}/P$ ratio; possibly, no medium- or low-income country can achieve a high $P_{top\ 10\%}/P$ ratio. However, among countries with similar GDP per capita and technological activity, the $P_{top\ 10\%}/P$ ratio can vary widely depending on the research policy.

**Funding**

This work was supported by the Spanish Ministerio de Ciencia e Innovación [grant number PID2020-113455GB-I00]

**Supplementary data** can be requested to the authors